\documentclass[superscriptaddress, aps, pra, onecolumn, nofootinbib]{revtex4-2}

\usepackage{amsfonts}
\usepackage{amsmath}
\usepackage{graphicx}% Include figure files
\usepackage{dcolumn}% Align table columns on decimal point
\usepackage{bm}% bold math
\usepackage[colorlinks, allcolors={blue}]{hyperref}% add hypertext capabilities

\begin{document}

\title{Aharonov-Bohm effect with an effective complex-valued vector potential}

\author{Ismael L. Paiva}%\orcid{0000-0002-0416-3582}}
\email{ismaellpaiva@gmail.com}
\affiliation{H. H. Wills Physics Laboratory, University of Bristol, Tyndall Avenue, Bristol BS8 1TL, United Kingdom}
\affiliation{Schmid College of Science and Technology, Chapman University, Orange, California 92866, USA}
\affiliation{Institute for Quantum Studies, Chapman University, Orange, California 92866, USA}

\author{Yakir Aharonov}
\affiliation{Schmid College of Science and Technology, Chapman University, Orange, California 92866, USA}
\affiliation{Institute for Quantum Studies, Chapman University, Orange, California 92866, USA}
\affiliation{School of Physics and Astronomy, Tel Aviv University, Tel Aviv 6997801, Israel}

\author{Jeff Tollaksen}
\affiliation{Schmid College of Science and Technology, Chapman University, Orange, California 92866, USA}
\affiliation{Institute for Quantum Studies, Chapman University, Orange, California 92866, USA}

\author{Mordecai Waegell}
\affiliation{Schmid College of Science and Technology, Chapman University, Orange, California 92866, USA}
\affiliation{Institute for Quantum Studies, Chapman University, Orange, California 92866, USA}

\begin{abstract}
The interaction between a quantum charge and a dynamic source of a magnetic field is considered in the Aharonov-Bohm scenario. It is shown that, in weak interactions with a post-selection of the source, the effective vector potential is, generally, complex-valued. This leads to new experimental protocols to detect the Aharonov-Bohm phase before the source is fully encircled. While this does not necessarily change the nonlocal status of the Aharonov-Bohm effect, it brings new insights into it. Moreover, we discuss how these results might have consequences for the correspondence principle, making complex vector potentials relevant to the study of classical systems.
\end{abstract}

\maketitle

\section{Introduction}

The Aharonov-Bohm (AB) effect~\cite{ehrenberg1949refractive, Aharonov1959, aharonov1963further} refers to the relative phase $\phi_{AB} = q\Phi_B/\hbar$ acquired by a quantum particle with charge $q$ that encircles, but does not enter, a region with magnetic flux $\Phi_B$ on its interior. In this scenario, there is a sense in which the charge interacts with the vector potential associated with the magnetic field, which is always non-zero in at least part of the particle's trajectory. The description of such an interaction before the interference is gauge-dependent. However, these descriptions are regarded as nonphysical since they do not lead to observable effects. Nevertheless, the charge's final state, i.e., the state with a phase shift characterized by $\phi_{AB}$ has a clear physical meaning. This phenomenon has been applied in many areas~\cite{Olariu1985, Peshkin1989, berry1989quantum, ford1994aharonov, vidal1998aharonov, tonomura2006aharonov, recher2007aharonov, russo2008observation, peng2010aharonov, fang2012photonic, bardarson2013quantum, noguchi2014aharonov, duca2015aharonov, mukherjee2018experimental, paiva2019topological, cohen2019geometric, paiva2020magnetic, paiva2022geometric}.

One characteristic of the standard AB effect is that the source of the magnetic field or even the magnetic field itself is not considered to be dynamical, i.e., they are taken to be parameters that affect the dynamics of the charge encircling the region of interest. However, the analysis of the effect becomes more subtle when the source of the magnetic field is quantized~\cite{peshkin1961quantum, aharonov1991there, santos1999microscopic, choi2004exact, Vaidman2012, pearle2017quantum, pearle2017quantized, li2018transition, marletto2020aharonov, horvat2020probing, saldanha2021shielded}. Of particular interest here is the model studied in Ref.~\cite{aharonov1991there}, where an entanglement between the source and the encircling charge is manifest as a consequence of the AB effect. Specifically, this model considers an infinitely long cylindrical shell, with a moment of inertia $I_c$, and uniform charge density rotating around its long axis of symmetry (say, the $z$ axis) with angular speed $\dot{\varphi}$, as represented in Fig.~\ref{fig:AB}. Also, a uniformly charged wire is added in the $z$ axis to cancel the electric field on the exterior of the cylinder. Thus, in this configuration, the cylinder is analogous to a solenoid. The AB effect can, then, be studied by considering a charge $q$ with mass $m$ and moment of inertia $I_q$ encircling the cylinder. The charge can be assumed to move in the $xy$ plane and have polar coordinates $r$ and $\theta$. As shown in Ref.~\cite{aharonov1991there} and briefly reviewed in Appendix~\ref{app}, neglecting radiation terms, the Hamiltonian of the joint system can be written as
\begin{equation}
    H = \frac{1}{2m} p_r^2 + \frac{1}{2I_q} \left(p_\theta - \frac{qK}{2\pi I_c} p_\varphi\right)^2 + \frac{1}{2I_c} p_\varphi^2,
    \label{eq:hamilt-class}
\end{equation}
where $K$ is a constant and $p_r$, $p_\theta$, and $p_\varphi$ are, respectively, the momentum associated with $r$, $\theta$, and $\varphi$.

The quantization of this Hamiltonian will be the basis for our result. Observe that the expression in Eq.~\eqref{eq:hamilt-class} is written in the Coulomb (or Lorenz) gauge. Indeed, this can be observed not only by the algebraic form of the vector potential-like term $qK p_\varphi/2\pi I_c$ but also by the fact that $p_\varphi$ is divergence free. Then, one may ask if, in our study and other related ones, a privileged status is given to this gauge. The goal of Ref.~\cite{aharonov1991there} was to answer this question. There, it was pointed out that a non-trivial change of gauge leads to a Hamiltonian that satisfies a constraint. By solving the constraint at the classical level, the authors concluded that, although the resultant Hamiltonian is given by a different expression in the new gauge, it is so because of a different choice of coordinates, i.e., a different reference frame. In this new frame, the coordinate $\theta$ remains unchanged. However, $\varphi$ is transformed to a linear combination of $\varphi$ and $\theta$. Therefore, in our study, the coordinate system is fixed in such a way that the Hamiltonian of the system is of the form in Eq.~\eqref{eq:hamilt-class}.

The objective of this article is to study the joint system in scenarios with post-selections in the state of the cylinder and to show how the inclusion of the flux source on the dynamics modifies the AB effect, leading to new predictions and even a manifestation of the AB effect before the charge completes its loop. These consequences arise from the emergence of a complex-valued effective vector potential associated with the magnetic field source. Furthermore, we discuss how this analysis brings new insights into the correspondence principle.

\section{Results}

\subsection{Dynamics of the joint system}

The initial motivation for our search comes from the analysis of the classical dynamics of the joint system, given by Hamilton's equations, which imply that
\begin{equation}
    \dot{\varphi} = \frac{\partial H}{\partial p_\varphi} = - \frac{qK}{2\pi I_c I_q} \left(p_\theta - \frac{qK}{2\pi I_c} p_\varphi\right) + \frac{1}{I_c} p_\varphi
\end{equation}
and
\begin{equation}
    \dot{\theta} = \frac{\partial H}{\partial p_\theta} = \frac{1}{I_q} \left(p_\theta - \frac{qK}{2\pi I_c} p_\varphi\right).
\end{equation}
As a result,
\begin{equation}
    \dot{\varphi} = - \frac{qK}{2\pi I_c} \dot{\theta} + \frac{1}{I_c} p_\varphi.
\end{equation}
Hence, a change in the charge's angular position modifies the angular speed of the cylinder. This correlation will also have an effect on the quantum treatment of the problem.

In fact, after canonical quantization, the Hamiltonian in Eq.~\eqref{eq:hamilt-class} becomes \cite{aharonov1991there}
\begin{equation}
    H = \frac{1}{2\mu} P_r^2 + \frac{1}{2I_q} \left(P_\theta - \frac{qK}{2\pi I_c} L_z\right)^2 + \frac{1}{2I_c} L_z,
    \label{eq-hamilt}
\end{equation}
where $P_r$ and $P_\theta$ are, respectively, the canonical radial and angular momentum operators of the charge, and $L_z$ is the canonical angular momentum operator of the cylinder. Observe that, with a quantized source, the Hamiltonian does not contain a vector potential \textit{per se}. Instead, it contains an ``operator vector potential'' $\vec{A} = (K/2\pi I_c r) L_z \hat{\theta}$.

\begin{center}
\begin{figure}
\includegraphics[width=.6\columnwidth]{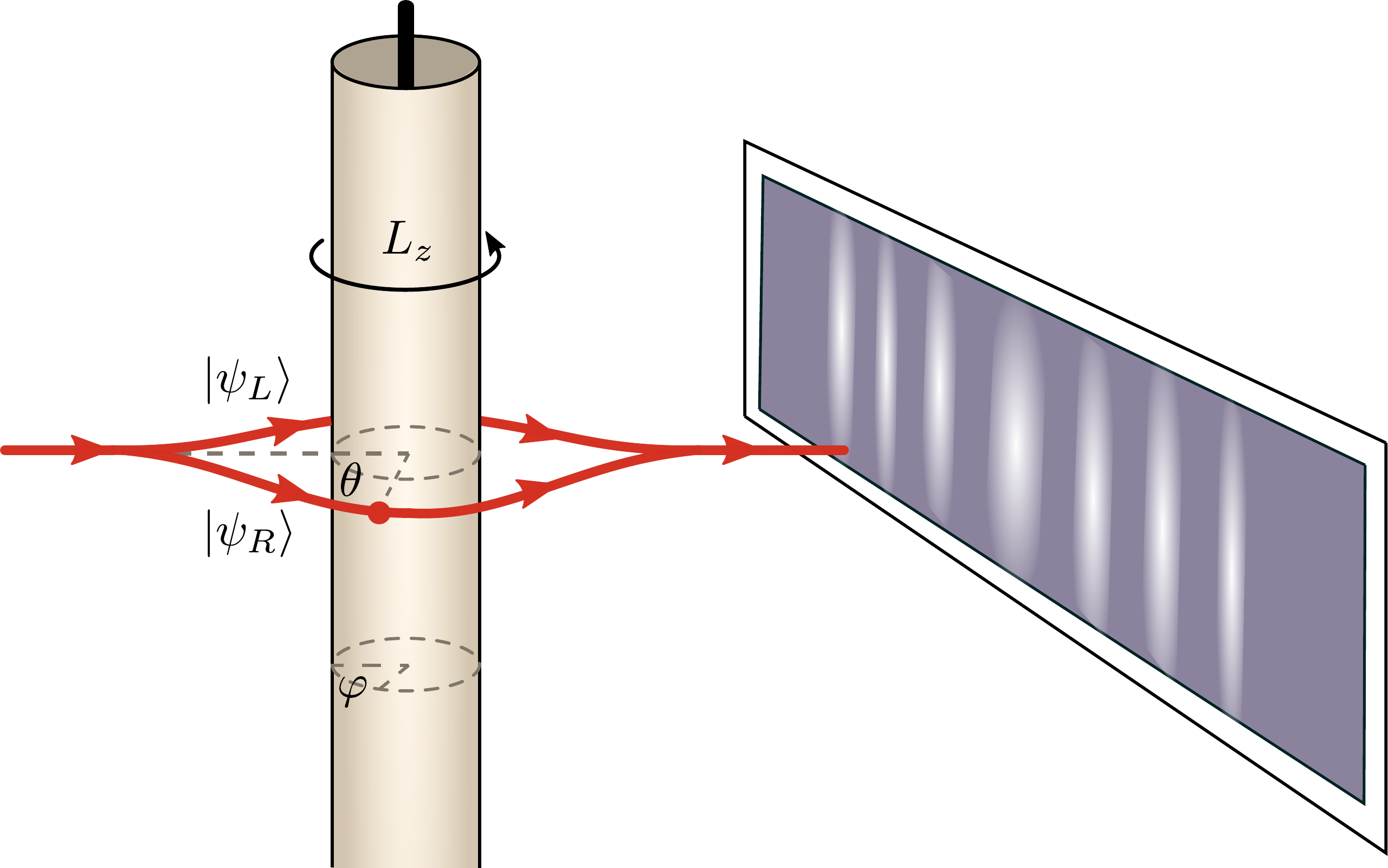}
\caption{\textbf{Aharonov-Bohm effect with a dynamic magnetic field source.} An infinitely long cylindrical shell of charges, when rotating, produces a magnetic field proportional to its angular momentum $L_z$ on its interior. Outside, it produces an electric field, which is canceled by the electric field of a uniform line of charge (thick vertical line). In this scenario, a quantum charge encircling the cylinder in a superposition of $|\psi_L\rangle$ and $|\psi_R\rangle$ (curved red trajectories) acquires a quantum phase that can be observed in the interference pattern on a screen. If the cylinder is in a superposition of states with definite $L_z$ the interference pattern is generally destroyed.}
\label{fig:AB}
\end{figure}
\end{center}

We are interested in investigating the AB scenario, as illustrated in Fig.~\ref{fig:AB}. For this, let the state of a quantum charge as it begins to enclose the cylinder be given by
\begin{equation}
    |\psi_0\rangle = \frac{1}{\sqrt{2}} \left(|\psi_L\rangle+|\psi_R\rangle\right),
    \label{eq:initial-state}
\end{equation}
where $|\psi_L\rangle$ represents a packet that passes to the left of the cylinder and $|\psi_R\rangle$ a packet that travels to the right of it. Also, denote by $|m_\ell\rangle$ an eigenstate of $L_z$ with eigenvalue $m_\ell\in\sigma_\ell \equiv \{-\ell, -\ell+1, \cdots, -1, 0, 1, \cdots, \ell-1, \ell\}$ and $\ell$ is a natural number. Finally, let 
\begin{equation}
    |\phi_0\rangle = \sum_{m_\ell\in\sigma_\ell} c_{m_\ell} |m_\ell\rangle.
    \label{am-state}
\end{equation}
be the initial state of the cylinder, where $c_{m_\ell}\in\mathbb{C}$ such that $\sum_{m_\ell\in\sigma_\ell} |c_{m_\ell}|^2 = 1$. Then, the state of the joint system while the charge is inside the AB loop is
\begin{equation}
    |\Lambda(\theta)\rangle = \sum_{m_\ell\in\sigma_\ell} \frac{c_{m_\ell}}{\sqrt{2}} \left(e^{-i qKm_\ell \theta/2\pi I_c} |\psi_L\rangle + e^{iKqm_\ell \theta/2\pi I_c} |\psi_R\rangle\right) \otimes |m_\ell\rangle,
    \label{map}
\end{equation}
where $\theta$ refers to the angular ``distance'' travelled by each packet. In the above expression, for simplicity, phases associated with the terms $P_r^2/2\mu$ and $L_z^2/2I_c$ were omitted since they are not relevant to the results of interest in this work.

Observe that the moving charge and the cylinder become entangled. Moreover, the state of the charge alone $\rho = \text{Tr}_\text{cylinder} \left(|\Lambda(\theta)\rangle \langle\Lambda(\theta)|\right)$ is
\begin{equation}
    \rho = \frac{1}{2} \left[|\psi_L\rangle\langle\psi_L|+|\psi_R\rangle\langle\psi_R| + \left(\sum_{m_\ell\in\sigma_\ell} \left|c_{m_\ell}\right|^2 e^{iqKm_\ell \theta/\pi I_c}\right) |\psi_R\rangle\langle\psi_L| + \left(\sum_{m_\ell\in\sigma_\ell} \left|c_{m_\ell}\right|^2 e^{-iqKm_\ell \theta/\pi I_c}\right) |\psi_L\rangle\langle\psi_R|\right].
    \label{eq:dmatrix}
\end{equation}

\subsection{Effective complex-valued vector potentials}

If $I_c \gg Kqm_\ell$ for every $m_\ell$ such that $c_{m_\ell} \neq 0$, it follows that
\begin{equation}
    \begin{aligned}
        \sum_{m_\ell\in\sigma_\ell} \left|c_{m_\ell}\right|^2 e^{iqKm_\ell \theta/\pi I_c} &= \sum_{m_\ell\in\sigma_\ell} \left|c_{m_\ell}\right|^2 \left[1 + \frac{iqKm_\ell \theta}{\pi I_c} + O\left(\frac{1}{I_c^2}\right) \right] \\
        &= 1 + \sum_{m_\ell\in\sigma_\ell} \left|c_{m_\ell}\right|^2 \frac{iqKm_\ell \theta}{\pi I_c} + O\left(\frac{1}{I_c^2}\right) \\
        &\approx e^{iqK \left(\sum_{m_\ell\in\sigma_\ell} \left|c_{m_\ell}\right|^2 m_\ell\right) \theta/\pi I_c} = e^{iqK \langle L_z\rangle \theta/\pi \hbar I_c}.
    \end{aligned}
    \label{eq:approx}
\end{equation}
Observe that this approximation approaches equality in the limit $I_c \rightarrow \infty$. Indeed, the expressions on the left-hand side and on the last line of the right-hand side are equivalent to the first order in $1/I_c$.

The approximation in Eq. \eqref{eq:approx} implies that the reduced density matrix in Eq. \eqref{eq:dmatrix} becomes a projector onto the state
\begin{equation}
    \frac{1}{\sqrt{2}} \left(|\psi_L\rangle+e^{iqK \langle L_z\rangle \theta/\pi\hbar I_c}|\psi_R\rangle\right)
    \label{eq:approx-pre},
\end{equation}
which is equivalent to having the cylinder effectively represented by the vector potential $\vec{A}(r) = (K \langle L_z\rangle/2\pi I_c r) \hat{\theta}$.

As a consequence of the quantization in other gauges already mentioned, the entanglement between the moving charge and the cylinder is described differently in distinct gauges. These alternative descriptions correspond to different choices of reference frames. This feature is consistent with results showing that superposition and entanglement are frame-dependent~\cite{aharonov1984quantum, angelo2011physics, giacomini2019quantum}.

Now, if the charge is stopped at a location $\theta<\pi$, no measurement will allow the observance of the AB effect. However, as we discuss next, this conclusion changes if a post-selection of the cylinder's state is considered.

Let the cylinder be post-selected in the state
\begin{equation}
    |\phi_1\rangle = \sum_{m_\ell\in\sigma_\ell} d_{m_\ell} |m_\ell\rangle.
    \label{post-selection}
\end{equation}
The state of the charge stopped at $\theta$ can be described by
\begin{equation}
    \begin{aligned}
        \langle\phi_1|\Lambda(\theta)\rangle & = \sum_{m_\ell\in\sigma_\ell} \frac{c_{m_\ell}d^*_{m_\ell}}{\sqrt{2}} (e^{-i qKm_\ell \theta/2\pi I_c} |\psi_L\rangle + e^{iKqm_\ell \theta/2\pi I_c} |\psi_R\rangle) \\
        &\approx \frac{\langle\phi_1| \phi_0\rangle}{\sqrt{2}} \left(|\psi_L\rangle+e^{iqK \langle L_z\rangle_w \theta/\pi\hbar I_c}|\psi_R\rangle\right)
    \end{aligned}
\end{equation}
if $I_c \gg Kqm_\ell$ for every $m_\ell$ such that $c_{m_\ell}d^*_{m_\ell} \neq 0$. In this expression, $\langle L_z\rangle_w \equiv \langle\phi_1| L_z |\phi_0\rangle/\langle\phi_1|\phi_0\rangle$ is the \textit{weak value}~\cite{aharonov1988result} of $L_z$. Observe that now the effective vector potential is $\vec{A}(r) = (K \langle L_z\rangle_w/2\pi I_c r) \hat{\theta}$.

This is our main result. What makes it noteworthy is the fact that weak values may lie outside the spectrum of the operators and, in particular, may be complex-valued. In fact, the normalized state that can be associated with the charge is
\begin{equation}
    \frac{1}{\sqrt{e^{\beta \theta} + e^{-\beta \theta}}} \left(e^{\beta \theta/2}|\psi_L\rangle + e^{-\beta \theta/2 + iqK \text{Re}(\langle L_z\rangle_w) \theta/\pi\hbar I_c}|\psi_R\rangle\right),
    \label{eq:state-middle}
\end{equation}
where $\beta=Kq\text{Im}\left(\langle L_z\rangle_w\right)/\pi \hbar I_c$ is a real number.

To see that it is indeed possible to design scenarios where the weak value of $L_z$---and hence the effective vector potential---has a complex value, assume the cylinder is prepared in a state that can be approximated by
\begin{equation}
    \phi_0(\varphi) = \frac{1}{(2\pi \Delta^2)^{1/4}} e^{-(\varphi-\alpha)^2/4\Delta^2},
\end{equation}
where $\alpha$ and $\Delta$ are positive real constants. Also, let the post-selection of the cylinder's state be
\begin{equation}
    \phi_1(\varphi) = \frac{1}{(2\pi \Delta^2)^{1/4}} e^{-(\varphi+\alpha)^2/4\Delta^2}.
\end{equation}
Then, since $L_z = -i\hbar \partial/\partial\varphi$, it follows that
\begin{equation}
    \begin{aligned}
        \langle\phi_1|L_z|\phi_0\rangle &= \frac{1}{(2\pi \Delta^2)^{1/2}} \int d\varphi \; e^{-(\varphi+\alpha)^2/4\Delta^2} \frac{i\hbar(\varphi-\alpha)}{2\Delta^2} e^{-(\varphi-\alpha)^2/4\Delta^2}\\
        &= -\frac{i\hbar\alpha}{2\Delta^2} \langle\phi_1|\phi_0\rangle,
    \end{aligned}
    \label{eq:wv}
\end{equation}
which implies that $\langle L_z\rangle_w = -i\hbar\alpha/2\Delta^2$, i.e., the weak value of $L_z$ is imaginary. Moreover, observe that one can take $\Delta\ll1$, making the imaginary component of $\langle L_z\rangle_w$ arbitrarily large.

These complex-valued vector potentials have many consequences, including the existence of AB-related effects before the charge completes its loop and an implication to the correspondence principle. The remainder of this article is dedicated to the study of these and other consequences.

\subsection{Observable effects of complex vector potentials}

We present now three observable consequences of effective complex-valued vector potentials, two of which do not require the enclosing of the AB loop.

To start, assume the charge completes the AB loop and, following that, the cylinder is post-selected in the state given by Eq.~\eqref{post-selection}. Then, the state $|\psi_L\rangle+e^{iqK \langle L_z\rangle_w/\hbar I_c}|\psi_R\rangle$ can be assigned to the charge. If $\text{Im}(\langle L_z\rangle_w)$ is non-null, the initial distribution between the left and right arms gets modified.

\begin{figure}
    \centering
    \includegraphics[width=.5\columnwidth]{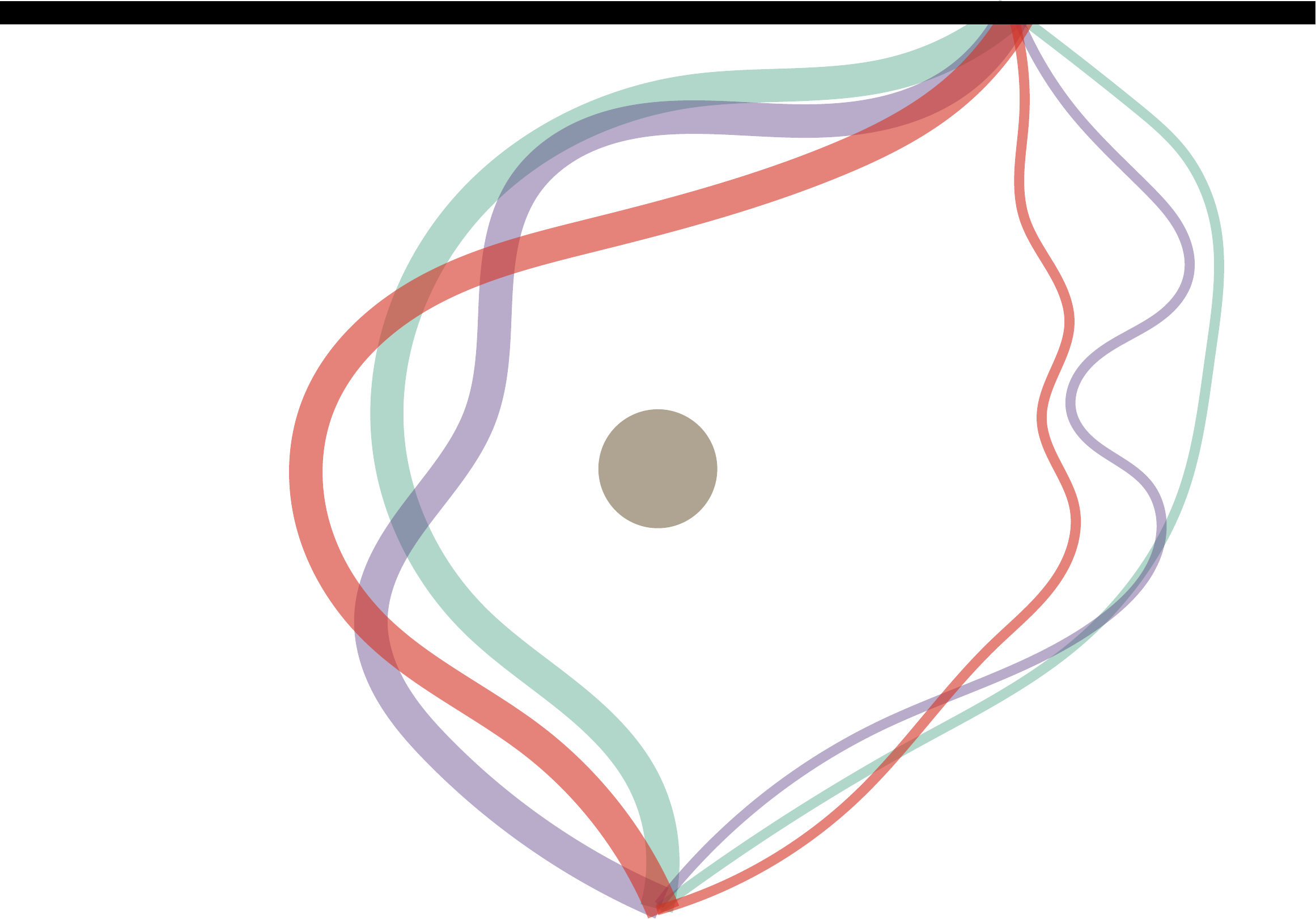}
    \caption{\textbf{Detection of an electron after it encircles a rotating charged cylinder with an associated complex-valued vector potential.} The cylinder is represented by the disk and the detecting screen, by the horizontal line. Even if the initial state of the electron corresponds to an equal distribution between the left and the right arms, the probability of finding the electron at a certain location equates to a scenario where the amplitudes on each arm differ. This result is independent of the shape of the trajectories, which evidences the topological nature of the effect.}
    \label{fig:amp}
\end{figure}

Then, if the charge is prepared in an even distribution of packets on both sides and is measured on a screen after completing the loop, the probability of finding it at a certain location is associated with different amplitudes traveling on each arm. This is the case regardless of the trajectory on each arm, as illustrated in Fig.~\ref{fig:amp}.

\begin{figure}
    \centering
    \includegraphics[width=.6\columnwidth]{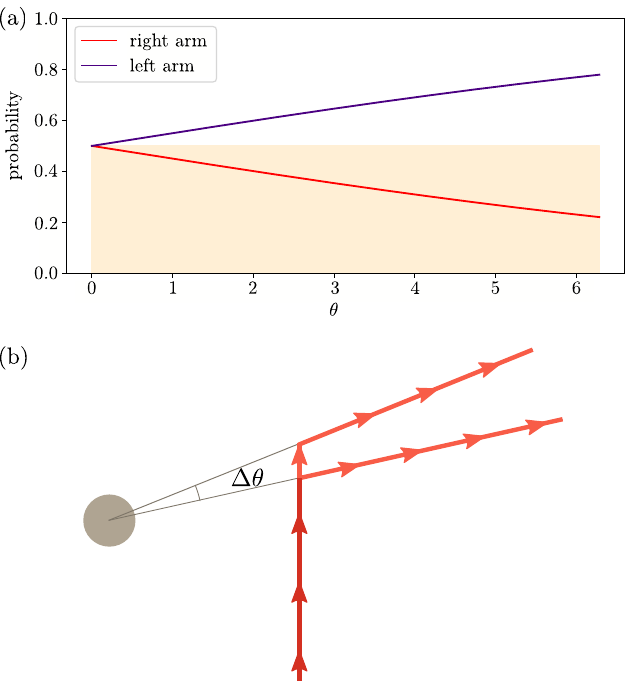}
    \caption{\textbf{Two examples of consequences of effective complex-valued vector potentials.} (a) It was assumed that the parameter $\beta$ in Eqs.~\eqref{pL} and \eqref{pR}, which is proportional to the imaginary term of the effective vector potential is $0.1$. The graph is associated with the AB effect for a given pre-selected cylinder state and an equal superposition of packets on each arm. Detectors are placed on each arm at an angle $\theta$. In this measurement, the charge will be found in either arm in half of the trials, as represented by the shaded area. If, however, these measurements are conditioned upon a subsequent post-selection of the cylinder, the proportion on each arm is altered, as shown by the purple and red curves. (b) The illustration shows a different manner to detect the complex-valued vector potential without the electron closing a loop around the cylinder (disk). At a certain location, the electron is evenly superposed. One of the packets travels radially while the other moves to a location associated with a different angle and, then, travels radially. Measurements of the electron followed by an appropriate post-selection of the cylinder should indicate different amplitudes on each arm. Even if the electron is not stopped during the measurement.}
    \label{fig:exp}
\end{figure}

The first observable effect before the completion of the AB loop we present corresponds to a change in amplitude of a moving charge encircling a magnetic field source and observed after each packet has traveled an angle $\theta$. To be more precise, detectors are placed at the locations $\theta$ and $-\theta$, where the charge is stopped in the state given by Eq.~\eqref{eq:dmatrix}. Following this, the cylinder is post-selected in the state in Eq.~\eqref{post-selection}, which assigns to the charge the state in Eq.~\eqref{eq:state-middle}. If this procedure is repeated many times, it will be observed that, before considering the post-selection, each detector clicks half of the time. However, after discarding the cases for which the result of the post-selection measurement differs from $|\phi_1\rangle$, it will be concluded that, in the remaining cases, the detector on the left arm fires with probability $p_L$ given by
\begin{equation}
    p_L(\theta) = \frac{e^{\beta\theta}}{e^{\beta\theta}+e^{-\beta\theta}}
    \label{pL}
\end{equation}
and the probability of finding the particle on the right arm is
\begin{equation}
    p_R(\theta) = \frac{e^{-\beta\theta}}{e^{\beta\theta}+e^{-\beta\theta}}.
    \label{pR}
\end{equation}
This is exemplified in Fig.~\ref{fig:exp}(a).

Note that there is a comparison that can be made between this result and the behavior of open quantum systems in which the Berry phases become complex~\cite{garrison1988complex, berry1990geometric, zwanziger1991measuring, kepler1991geometric, ning1992geometrical, bliokh1999appearance, carollo2003geometric, dietz2011exceptional, cohen2019geometric}. Of course, the system analyzed here is not open, and post-selections are not considered in the study of open quantum systems. However, as seen in Eqs.~\eqref{pL} and \eqref{pR}, the emergence of an effective complex-valued vector potential makes the evolution of the packet on each arm comparable to the evolution of open systems. This suggests, then, that appropriate choices of pre and post-selections in the scenario investigated here can be used to emulate the dynamics of open systems statistically.

Finally, another similar effect can be observed in an experiment where, after the electron is evenly superposed, one of the packets, say $|\psi_{\theta_0}\rangle$, travels in the radial direction. Meanwhile, the other packet, say $|\psi_{\theta_1}\rangle$, first moves to a different $\theta$-coordinate, after which it also travels exclusively in the radial direction. This scenario is represented in Fig.~\ref{fig:exp}(b). Once both packets move radially, the electron is measured without necessarily stopping it. Finally, the cylinder is post-selected in the state~\eqref{post-selection}. In this scenario, the observed amplitude on the arm of $|\psi_{\theta_0}\rangle$ will be
\begin{equation}
    p_{\theta_0} = \left(1+ e^{-\beta \Delta\theta}\right)^{-1},
\end{equation}
which is in agreement with the following state that can be assigned to the electron
\begin{equation}
    |\psi\rangle = F \left(|\psi_{\theta_0}\rangle+e^{iKq\langle L_z\rangle_w \Delta\theta/2\pi\hbar}|\psi_{\theta_1}\rangle\right),
\end{equation}
where $\Delta\theta$ is the angle difference associated with the radial trajectories and $F = 1/ \sqrt{1+ e^{-\beta \Delta\theta}}$ is a normalization constant.

\section{Discussion}

In this work, we have studied the AB effect with a quantized magnetic field source. It was observed that when the source is post-selected, a complex-valued vector potential may emerge. Furthermore, we presented observable consequences of this result, including two thought experiments that lead to observable local effects before the charge completes its loop around the source. These thought experiments, although challenging, may, in principle, be implemented in current devices. For instance, a proof-of-principle experiment might be possible in superconducting quantum interference devices, where the rotating Cooper pairs can be seen as an analog to the rotating cylinder.

We note that Ref.~\cite{horvat2020probing} analyzed measurements to probe the relative phase within the AB loop. The authors concluded that any local measurement on the charge with this aim would necessarily read only the gauge-invariant mechanical contribution to the phase. At first sight, this seems to challenge the conclusions we present here. However, observe that the scenario we consider is different. Indeed, we start with a measurement of the charge's location. This measurement alone does not reveal anything about the AB effect. Following this, we then post-select the magnetic field source and verify the effects associated with the effective complex-valued vector potential. Thus our protocol does not probe the AB phase in the standard sense of this phase. Also, we reiterate that the ``gauge dependency'' here is due to a choice of a system of coordinates, i.e., a reference frame, as shown in Ref.~\cite{aharonov1991there} and briefly discussed in Appendix~\ref{app}.

It is also noteworthy that imaginary vector potentials have also been discussed in optical scenarios. More precisely, it has been shown that gauge materials inside coupled waveguides emulate imaginary vector potentials for optical systems~\cite{descheemaeker2017optical}. In our treatment, however, the effective complex-valued vector potential emerges in a scenario with post-selection in a similar way that the standard vector potential used in the semi-classical treatment of the AB effect appears. In this sense, the effective vector potential obtained here truly corresponds to the standard notion of the vector potential (and not an emulation of it).

Furthermore, as already discussed, in the scenarios considered here, the cylinder can be taken to be a macroscopic object with the state of uncertain angular momentum given by Eq.~\eqref{am-state}. If its moment of inertia $I_c$ is sufficiently high, it is possible to know both the cylinder's position and angular velocity with a relatively high degree of certainty (since its angular velocity is given by the ratio $L_z/I_c$). As a result, such a cylinder qualifies as a quasi-classical object. Now, if a quasi-classical macroscopic cylinder is post-selected in a state $|\varphi\rangle$ such that the effective vector potential is complex-valued, the complex vector potential is also a quasi-classical object, with the caveat that the post-selections of macroscopic objects is not an easy task to be physically realized according to our current technological capabilities and understanding of quantum operations.

As a consequence, similarly to the discussions in Ref.~\cite{bender2010complex, aharonov2014unusual, cohen2017quantum, aharonov2022complex}, the results presented here seem to require modifications of the correspondence principle, which refers to the connection between classical and quantum systems in a particular limit. In fact, in the examples discussed and, in particular, in the case associated with the weak value in Eq.~\eqref{eq:wv}, it is possible to take the classical limit by taking $\ell\rightarrow\infty$. Even if $\langle L_z\rangle_w \rightarrow\infty$, it is possible to take the limit in such a way that $\langle L_z\rangle_w/I_c$ is constant.

The Ehrenfest theorem, which is also seen as a mathematical basis for the correspondence principle, accounts only for real expectation values playing a role in classical physics. However, the existence of effective vector potentials with arbitrarily large imaginary terms, as shown here, seems to urge changes to these ideas that take post-selections of quantum systems and the weak values associated with them as relevant elements in the quantum-to-classical transition.

\section*{Acknowledgments}

The authors would like to thank Eliahu Cohen and Lev Vaidman for their comments on a previous version of this manuscript. I.L.P. acknowledges financial support from the Science without Borders Program (CNPq/Brazil, Fund No. 234347/2014-7) and the ERC Advanced Grant FLQuant. Y.A. thanks the Federico and Elvia Faggin Foundation for support. This research was supported by the Fetzer-Franklin Fund of the John E. Fetzer Memorial Trust and by Grant No. FQXi-RFP-CPW-2006 from the Foundational Questions Institute and Fetzer Franklin Fund, a donor-advised fund of Silicon Valley Community Foundation.

\appendix

\section{Quantization of a system with a dynamical source of magnetic field}
\label{app}

Here, we briefly review the result and discussion presented in Ref.~\cite{aharonov1991there}. To start, consider a classical system with an infinitely long cylindrical shell with radius $a$, a moment of inertia $I_c$, and uniform charge density $\sigma$ rotating around its long axis of symmetry (say, the $z$ axis) with angular velocity $\dot{\varphi}$. It is known that such a cylinder generates a uniform magnetic field on its interior given by $\vec{B}=\mu_0\sigma\dot{\varphi} a \hat{z}$. Outside of it, the magnetic field vanishes. Moreover, assume a uniformly charged wire is added in the $z$ axis to cancel the electric field on the exterior of the cylinder. Also, let $(r,\theta)$ be the coordinates of a charge $q$ with mass $\mu$ traveling outside the cylinder in the $xy$ plane.

The Lagrangian that describes the dynamics of the joint system composed of the cylinder and the charge can be furnished as
\begin{equation}
    \mathcal{L} = \frac{1}{2}\mu\dot{r}^2 + \frac{1}{2}I_q\dot{\theta}^2 + \frac{1}{2}I_c\dot{\varphi}^2 + \frac{qK}{2\pi} \dot{\varphi} \dot{\theta},
    \label{eq-lag}
\end{equation}
where $I_q = \mu r^2$ is the moment of inertia of the charge and $K$ is a constant associated with the strength of the interaction between the cylinder and the particle.

It can be verified that
\begin{equation}
    \left\{\begin{array}{l}
        p_r = \mu \dot{r} \\
        p_\theta = I_q \dot{\theta} + \frac{qK}{2\pi} \dot{\varphi} \\
        p_\varphi = \frac{qK}{2\pi} \dot{\theta} + I_c \dot{\varphi}
    \end{array} \right..
\end{equation}
Then, neglecting terms of order $K^2$ and higher, which is valid whenever the coupling between the angular speeds of the source and the moving charge is weak, it follows that
\begin{equation}
    \begin{aligned}
        H &= p_r\dot{r} + p_\theta\dot{\theta} + p_\varphi \dot{\varphi} - \mathcal{L} \\
        &\approx \frac{1}{2\mu} p_r^2 + \frac{1}{2I_q} p_\theta^2 - \frac{qK}{2\pi I_c I_q} p_\theta p_\varphi + \frac{1}{2I_c} p_\varphi^2 \\
        &\approx \frac{1}{2\mu} p_r^2 + \frac{1}{2I_q} \left(p_\theta - \frac{qK}{2\pi I_c} p_\varphi\right)^2 + \frac{1}{2I_c} p_\varphi^2,
    \end{aligned}
    \label{eq:4}
\end{equation}
which is the Hamiltonian in Eq.~\eqref{eq:hamilt-class}.

By analogy with the standard case, where the source is not considered dynamical, observe that the vector potential associated with the cylinder is
\begin{equation}
    \vec{A} = \frac{K}{2\pi I_q r} p_\varphi \hat{\theta},
\end{equation}
i.e., it is consistent with the Coulomb or Lorenz gauge.

A question that arises from this approach, which has been addressed in Ref.~\cite{aharonov1991there}, concerns the quantization of this system in a different gauge. As presented there, the inclusion of the source introduces a term that depends on the angular acceleration of the source to the Lagrangian of the joint system in other gauges, which, in turn, leads to a change of the canonical coordinate of the cylinder. More specifically, the coordinate $\varphi$ is replaced by a new coordinate $\varphi'(\theta,\varphi)$. As a result, when the system is quantized in another gauge, it is quantized in a new system of coordinates, i.e., a different frame. Also, the new canonical operators do not obey the canonical commutations. However, the physical predictions of the interaction between the charge and the cylinder are, as they should be, the same in every gauge. The conclusion is that the Coulomb gauge, although not preferred, provides a more straightforward mathematical treatment of the problem.

For comparison, note that, in standard approaches to the AB effect, the variable associated with the cylinder ($\dot{\varphi}$) is not an independent dynamical variable in the Lagrangian. In fact, it is assumed to be an external parameter. Thus, the change of gauge can be introduced without the aforementioned change of the cylinder's coordinate since this coordinate is not dynamical and, then, does not impose any technical difficulty in obtaining the Hamiltonian from the Lagrangian.

\bibliography{citations}

\end{document}